# Improvement of Multiparametric MR Image Segmentation by Augmenting the Data with Generative Adversarial Networks for Glioma Patients

Eric Carver[1,2], Zhenzhen Dai[1], Evan Liang[1], James Snyder[1], Ning Wen[1*]

[1]Department of Radiation Oncology, Henry Ford Health System, Detroit, MI, USA
[2]Department of Oncology, Wayne State University, Detroit, MI, USA

**\* Correspondence:**
Dr. Ning Wen
nwen1@hfhs.org

**Keywords: GAN, U-Net, Glioma, GBM, Segmentation.**

Abstract

Every year thousands of patients are diagnosed with a glioma, a type of malignant brain tumor. Physicians use MR brain images as a key tool in the diagnosis and treatment of these patients. Neural networks show great potential to aid physicians in the medical image analysis. This study investigates the use of varying amounts of synthetic brain T1-weighted (T1), post-contrast T1-weighted (T1Gd), T2-weighted (T2), and T2 Fluid Attenuated Inversion Recovery (FLAIR) MR images created by a generative adversarial network to overcome the lack of annotated medical image data in training separate 2D U-Nets to segment enhancing tumor, peritumoral edema, and necrosis (non-enhancing tumor core) regions on gliomas. These synthetic MR images were assessed quantitively (SSIM=0.79) and qualitatively by a physician who found that the synthetic images seem stronger for delineation of structural boundaries but struggle more when gradient is significant, (e.g. edema signal in T2 modalities). Multiple 2D U-Nets were trained with original BraTS data and differing subsets of a quarter, half, three-quarters, and all synthetic MR images. There was not an obvious correlation between the improvement of values of the metrics in separate validation dataset for each structure and amount of synthetic data added, there is a strong correlation between the amount of synthetic data added and the number of best overall validation metrics. Out of the twelve validation metrics, one validation metric was highest of any reported values when a quarter of synthetic data was used during training, three were the highest when half the synthetic data was used, four were highest when three-quarters of the synthetic data was added, and six were the highest when all synthetic images were used during training. In summary, this study showed ability to generate high quality synthetic Flair, T2, T1, and T1CE MR images using the GAN. Using the synthetic MR images showed encouraging results to improve the U-Net segmentation performance which has the potential to address the scarcity of readily available medical images.

# 1.0 Introduction

Approximately 121,000 [1] people in the US are diagnosed with a malignant brain tumor annually, and more than 13,000 of those being Glioblastoma (GBM) WHO grade IV tumors with an unacceptable median overall survival despite best available treatment of less than to 2 years. For primary brain tumors WHO grade II-IV there are no curative treatments and limited approved therapies. Current management of primary brain tumors has two standard benchmarks, tissue analysis for diagnosis and the longitudinal analysis of treatment response/ tumor stability through serial brain tumor imaging. In fact, the brain MRI in patients with GBM is used to stratify clinical trial options prior to initial surgery and to offer patients definitive cytoreduction surgery for malignant glioma or GBM when radiographic features are highly suggestive of a malignant tumor. Therefore, advanced imaging methods to stratify patients into phenotypic, functional, molecular, and prognostic groups is highly sought after.

Amongst, GBM researchers, clinicians, patients, and patient advocates there is hope that new advances as promised by molecular targeted therapies, advanced radiation techniques, evolving surgical technologies, and unforeseen innovation will result in improved patient outcomes. Central to all of these is radiographic imaging; both for diagnosis and longitudinal patient monitoring. Applications of deep machine learning in brain tumor imaging has the potential to transition from a subjective analysis to objective analysis and create a new set of tools to refine treatment options, improve care quality, and ultimately impact patient care. One critical limitation to achieving the success seen in non-medical imaging is the volume of data needed to power deep machine learning. One possible solution to overcoming the limited brain tumor imaging data available for analysis is to create synthetic brain tumor MR images. Herein, we evaluate the performance of image segmentation using U-Net by augmenting the data with generative adversarial networks (GAN). Specifically, we investigate the accuracy of segmentation of whole tumor, enhancing tumor, and tumor core for glioma patients using T1-weighted (T1), post-contrast T1-weighted (T1Gd), T2-weighted (T2), and T2 Fluid Attenuated Inversion Recovery (FLAIR) images, and create an enhanced imaging data repository through creating synthetic brain tumor imaging.

The 2D U-Net [2] is the model used to create these segmentations for this study. It was chosen as it outperforms other models, such as the sliding window convolutional network, and is used in similar studies due to its ability to retain spatial information. It is common knowledge that deep learning techniques are highly powerful when the amount of training samples is large. However, in the medical field, especially in clinical trials, where limited numbers of training samples are accessible, deep learning models are easily overfitting during training stage and perform poorly in prediction. [3] Besides, annotation of medical images requires clinicians to be well-trained and experienced and is generally expensive and time-consuming.

Synthesizing new images as training samples provides a possible solution to overcome the challenge of limited number of annotated medical images. The original idea of synthesizing images indistinguishable from reality is inspired by the development of GANs. [4] Researchers then have leveraged GANs in a conditional setting which allows the model to deterministically control the generation of particular samples based on external information. [5-7] However, some researchers suggested that adversarial training might be unstable or even diverge, and introduced image per-pixel loss and perceptual loss. [8] In this paper, we proposed an Augmentation Network that was trained in a supervised fashion using paired brain maps and real patient MRI images and generated new training samples from manipulated brain maps.

## 2.0    Methods

### 2.1    Patient Population

Data was obtained from the BraTS multimodal Brain Tumor Segmentation Challenge 2018. [9-12] 19 different institutions provided a total of 210 patients for training and 66 patients for validation. T1, T1CE, T2, and FLAIR MR images were provided for each patient. BraTS provided accurate delineation of enhancing tumor, peritumoral edema, and necrosis (non-enhancing tumor core) performed by one to four clinicians and approved by neuro-oncologists.

### 2.2    Image Pre-Processing

T1, T2, and FLAIR MRI were rigidly registered with the T1CE, resampled ($1 \times 1 \times 1mm^3$), skull stripped, and normalized. The enhancing tumor, peritumoral edema, and necrosis (non-enhancing tumor core) regions are small compared to the total 3D MRI volumes (256×256×155) creating a data imbalance. To combat this problem, 64 slices image patches covering these regions were extracted. Data augmentation was done by flipping each slice left/right to decrease dependence on location as the brain exhibits marked symmetry across the sagittal plane. All 155 slices of each MRI 3D volume were segmented during validation.

### 2.3    Generative Adversarial Neural Network

To overcome the limitation imposed by limited volume of MR data, the deep learning model named as Augmentation Network was employed to create artificial MR images from real images by changing tumor size, location and orientation. The Augmentation Network works by creating two competing networks. The generator creates the MR images, while the discriminator judges the quality of the MR image created by the generator. If the discriminator decides the generator created image lacks the quality of real images used for training, the generator is penalized. This process repeats as the generator changes parameters until the image quality is passed by the generator.

### 2.3.1 Architecture

Our augmentation network consists of a generator (blue box in Figure 1) and two discriminators (red and yellow box in Figure 1). The generator is used to generate synthetic MRI images from brain maps which are derived from real MRI images. The brain map is composed with the segmentation of normal brain tissue and defined sub-regions of GBM (enhancing tumor, necrosis/non-enhancing tumor and edema). The segmentation of normal brain tissue is achieved by image thresholding (3/4, 1/2 and 1/4 of max pixel value). The discriminators are used to distinguish between synthetic images and real patient images.

### 2.3.2 Generator

The generator consists of several components $C_i$, each operates at a different resolution. The brain map (*256 × 256*) is down-sampled to provide segmentation layout at different resolutions (*wi × hi, wi = hi*). The first component *C0* gets down-sampled brain map at the resolution of *w0 = h0 = 4* as input and generates feature maps as output to the next component. For components *C1* to *Cn*, feature maps from previous component are up-sampled at scale of 2 and are concatenated with brain map of the same resolution as input. Residual block is applied to generate feature maps as output. Convolution kernel size is *3×3*, layer normalization[13] is applied, LReLU [14] is used as the activation function.

### 2.3.3 Discriminator

(3) Two discriminators are used. The first one is a pre-trained VGG-19 convolutional neural network, [15] which won the first and second place in localization and classification in the Image Net Large Scale Visual Recognition Challenge (ILSVRC) 2014. It is used to calculate the perceptual loss ($\sum_i p_i$) and the image per-pixel loss $\mathcal{L}_{im}$.

$$\mathcal{L}_{im} = \sum_m \sum_n ||I_{real} - I_{synthetic}||_1 \quad (1)$$

$I_{real}$ represents for real patient MRI image, $I_{synthetic}$ represents for synthetic MRI image from the generator.

$$p_i = \sum_m \sum_n ||\theta_{i_{real}} - \theta_{i_{synthetic}}||_1 \quad (2)$$

$p_i$ is the perceptual loss from layer *i* of VGG-19 Net. $\theta_{i_{real}}$ and $\theta_{i_{synthetic}}$ are feature maps of patient MRI image and synthetic MRI images generated at layer *i* respectively. The second one is a patch GAN, which penalizes on image patches, the loss is given as $\mathcal{L}_{adv} = \mathbb{E}[D(I_{real}, I_{synthetic})] + \mathbb{E}[1 - D(I_{synthetic})]$. $D(\cdot)$ is the discriminator net. Total loss is computed as the weighted summation of each loss.

The synthetic image is acquired by solve the following objective:

$$S^* = argmin(\mathbb{E}[\sum_{i=0}^{n} \lambda_i p_i, + \lambda_{im}\mathcal{L}_{im}] + \lambda\mathcal{L}_{adv}) \quad (3)$$

### 2.3.4 Training and generation of new training samples

164 patients were randomly selected from BraTS18 dataset for training. For each MRI modality, an independent model was trained. $\lambda_i, \lambda_{im}$ and $\lambda$ were adapted every 10 epochs to maintain balance among each loss. Total training epoch is 100 for each modality. Then the brain map was manipulated for different lesion location, shape or size. Generated images and manipulated brains maps were used as new training samples to train the new model.

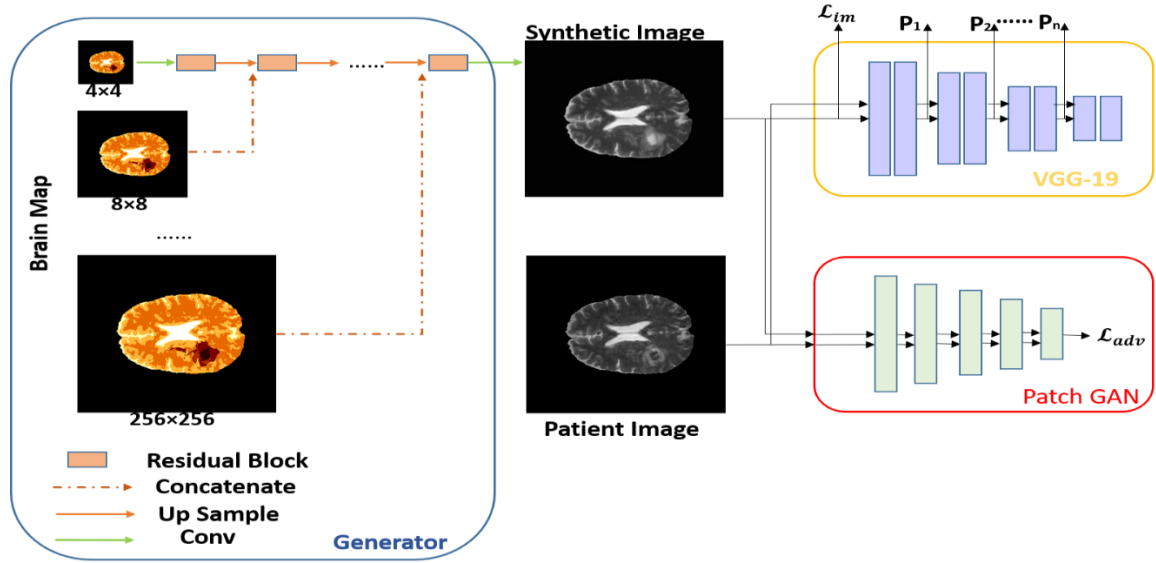

Figure 1. Architecture of Augmentation Net. Generator is shown in blue box and discriminators in yellow and red boxes.

## 2.4 U-Net

T1, T2, and FLAIR MRI used for input for three individual U-Nets. Whole tumor (WT), enhancing tumor (ET), and tumor core (TC) each received their own U-Net. The U-Net followed Pelt *et al.'s* [16] recommendation of four scaling layers. The U-Net, designed by Ronneberger, *et al.* [2] maintains the spatial information while doubling features for each consecutive convolutional layer with batch normalization [12] and halving the number of features for each consecutive up-sampling layer. These layers combined to form a merged layer and soft dice (equation 4) was employed as the loss function.

$$\text{Dice Loss} = \frac{2*<y_{true},y_{pred}>+c}{<y_{true},y_{true}>+<y_{pred},y_{pred}>+c} \quad (4)$$

$y_{true}$ is clinician's contour, $y_{pred}$ is model's output, and $c$ (0.01) is a constant to avoid division-by-zero singularities.

Each U-Net was able to produce the corresponding mask they were trained on creating a whole tumor model, an enhancing tumor model, and a tumor center model. To improve ET/TC accuracy, the intensity difference between T1CE and T1 was calculated on a pixel level in each slice and only ET/TC predictions within the whole tumor contour were accepted. The best model for each type of contour was chosen according to the validation loss within 100 epochs run on GPU (Titan XP, nVidia, Santa Clara, CA).

## 2.5 Statistics

The metrics used to evaluate the agreement between the created U-Net model and the given reference contours are the dice similarity coefficient (DSC), Hausdorff distance (HD), sensitivity, and specificity.

Sensitivity (equation 5) is also known as a true positive, while specificity (equation 6) is known as a true negative.

$$Sensitivity = \frac{number\ of\ true\ positives}{number\ of\ true\ positives + number\ of\ false\ negatives} \qquad (5)$$

$$Specificity = \frac{number\ of\ true\ negativess}{number\ of\ true\ negatives + number\ of\ false\ postivies} \qquad (6)$$

The DSC (equation 7) gives statistic regarding general overlap:

$$\text{DSC} = \frac{2*<y_{true},\bar{y}_{pred}>}{<y_{true},y_{true}>+<\bar{y}_{pred},\bar{y}_{pred}>}, \qquad (7)$$

where $\bar{y}_{pred}$ is binary prediction using 0.5 as the threshold.

The Hausdorff distance with 95% confidence interval is the maximum distance of a point in one contour to the nearest point of the other contour:

$$h(A,B) = \max_{a \in A}\{\min_{b \in B}\{d(a,b)\}\} \qquad (8)$$

where *a* and *b* are points of sets *A* and *B*, respectively, and *d(a,b)* is Euclidean metric between these points[13].

Mean Square Error, Mean Absolute Error, Peak Signal to Noise Ratio, and Structural Similarity Index were used to quantitatively compare the synthetic MR images created by the generative adversarial network to the original MR images obtained from BraTS.

Mean Squared Error (MSE) is shown in equation 9, with 'n' being number of images being compared at once. Since MSE depends on intensity scaling, 16-bit images were used with pixel range 0-255.

$$MSE = \frac{1}{n}\sum\{(original\ image - generated\ image)^2\} \qquad (9)$$

Mean Absolute Error (MAE) is shown in equation 10 and determines the prediction error between the actual value and predicted value.

$$MAE = \frac{1}{n}\sum\{abs(original\ image\text{-}generated\ image)\} \qquad (10)$$

Peak Signal to Noise Ratio (PSNR) is shown in equation 11 and is reported in decibels (dB). This overcomes the limitation of MSE by scaling the MSE value according to image range, which is done by the $S^2$ value in eq 7. Generally, the higher the PSNR, the better the synthetic image; however, this may not always be the case. It is best to compare the PSNR of synthetic image with the same original image.

$$PSNR = -10 * log_{10}\left(\frac{MSE}{S^2}\right) \qquad (11)$$

Structural Similarity Index (SSIM) shows the perceived change in structural information as opposed to MSE, MAE, and PSNR that show absolute error differences. SSIM assumes pixels close to each other possess strong inter-dependency. It is based on luminance, contrast, and structure differences between the images and claimed to be the most accurate metric by many studies.

## 3.0 Results

### 3.1 Generative Adversarial Neural Network

We evaluated the synthetic MR images created by the GAN model qualitatively and quantitatively. Several metrics have been used for the quantitative analysis of synthetic versus original images. This study employed the commonly used mean square error, peak signal to noise ratio, mean absolute error, and structural similarity index. However, these metrics are influenced greatly as there is a major difference in tumor location between original and synthetic images as we purposely vary the tumor locations in the synthetic images. This creates complications for benchmark comparisons to other studies, as they do not change tumor location. There are additional limitations in these metrics as MSE depends on image pixel range (0-255 for this study) and PSNR is best used to compare different synthetic images to the same original image, it has less real meaning when compared between different datasets.

*Table One: Average reported Mean Square Error (MSE), Mean Absolute Error (MAE), Peak Signal to Noise Ratio (PSNR), and Structural Similarity Index (SSIM) for synthetic MR images created by GAN.*

|       | MSE          | MAE          | PSNR          | SSIM          |
|-------|--------------|--------------|---------------|---------------|
| T1    | 19.246±0.308 | 23.375±0.586 | 43.068±0.443  | 0.788 ± 0.002 |
| T1CE  | 19.249±0.274 | 22.805±0.583 | 43.054±0.437  | 0.789 ± 0.004 |
| T2    | 19.246±0.290 | 23.391±0.400 | 43.102±0.45   | 0.784 ± 0.003 |
| Flair | 18.930±0.40  | 24.119±1.48  | 43.126±0.46   | 0.794 ± 0.005 |

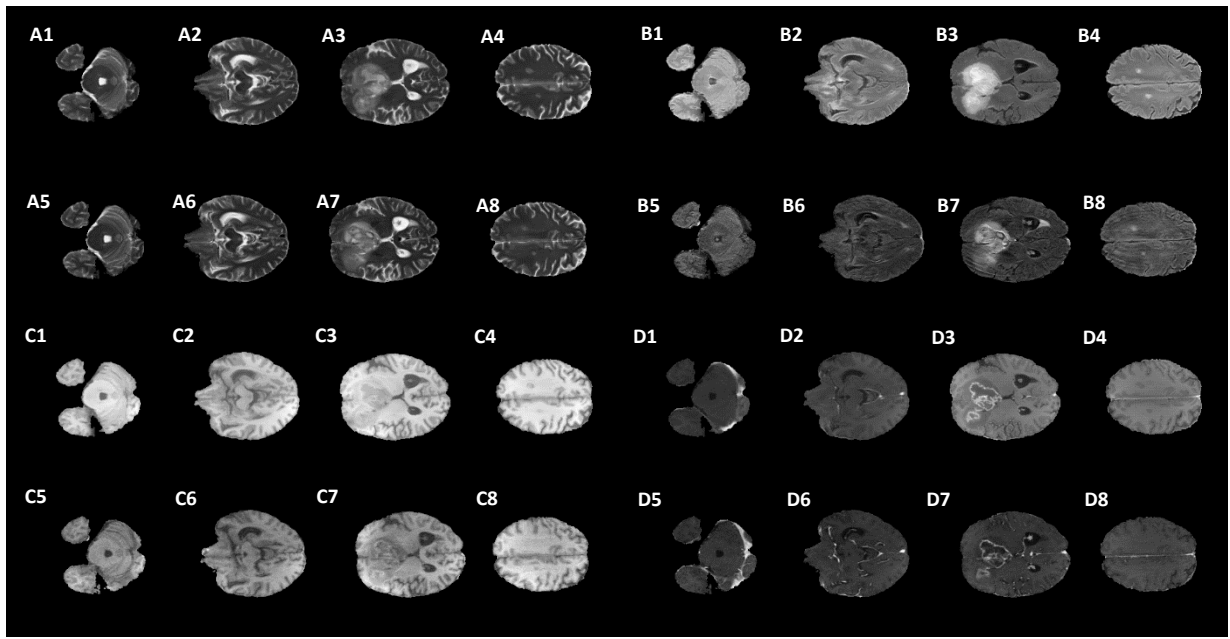

Figure 2: Synthetic MRI compared with patient MRI images (none manipulation). A1-A4: synthetic T2, A5-A8: patient T2; B1-B4: synthetic Flair, B5-B8: patient Flair; C1-C4: synthetic T1, C5-C8: patient T1; D1-D4: synthetic T1CE, D5-D8: patient T1CE.

In Figure 2, the difference between the synthetic image and the patient's actual image is shown in the image A3 vs. A7, specifically the tumor in the right frontal lobe (lower left on the image). The signal in the core of the tumor has increased T2 signal (brighter) compared with the surrounding edema. The geometric aspects of the tumor are preserved, but some of the relative signal intensities are distorted. Images A4 and A8 show signal from edema (fluid in brain). Image A4 has a wide range of contrast between areas of edema, whereas image A8 just delineates the extent of the edema with a sharp drop-off at the edges where that drop-off isn't physiologically that sharp. Images B5-B8 have some additional circumferential artifact. The extent of the edema on B7 vs B3 is also different. Synthetic quality for images C1-C4 (T1) are very good. T1 contrast images in D5-D8 are also geometrically good, although the area of decreased T1 signal surrounding the tumor in D3 seems to be less distinct on D7. In summary, the synthetic images seem stronger for delineation of structural boundaries but struggle more when gradient is significant (e.g. edema signal in T2 modalities).

In addition to these quantitative metrics, qualitative analysis for the generated MR images was assessed by a physician. To do this, a subset of 9 real images and 10 generated images flair, T1, T1CE, and T2 MR images were compared. The physician was presented with each of these 19 MR images and first judged if the MR image was real or synthetic, gave a score, and provided a comment. Ideally the real and synthetic images would be completely indistinguishable from each other, and this would be reflected by a 50 percent misclassification rate of the images. As shown in table three, Flair, T1CE, and T2 MR images were misclassified 26.3 percent of the time, while T1 was incorrectly identified as original or synthetic 10.5 percent of the time. This lower score was due to slight streaking being visible on coronal and sagittal views for some of the images.

images were compared. The physician was presented with each of these 19 MR images and first judged if the MR image was real or synthetic, gave a score, and provided a comment. Ideally the real and synthetic images would be completely indistinguishable from each other, and this would be reflected by a 50 percent misclassification rate of the images. As shown in table three, Flair, T1CE, and T2 MR images were misclassified 26.3 percent of the time, while T1 was incorrectly identified as original or synthetic 10.5 percent of the time. This lower score was due to slight streaking being visible on coronal and sagittal views for some of the images.

It should be noted that the 210 patients obtained from the BraTS competition ranged over many years. Since image quality has improved over these years, the newer images will always have better image quality than images generated from lower quality images from years before. Since all images were used in the generative adversarial network, the best this generative adversarial network could create would be the average of original image quality.

*Table Two: Physician review of subset of synthetic and original images. Percent misjudged shows the amount judged as original when it was synthetic and vice versa.*

| Modality | % Misclassified |
|---|---|
| Flair | 26.3 |
| T1 | 10.5 |
| T1CE | 26.3 |
| T2 | 26.3 |

**3.2 U-Net Independent Validation Results**

*Table Three: Validation Results for U-Nets trained by BraTS MRI and differing subsets of synthetic MRI. Underlined values show highest contour metric value. Bolded values indicate contour metric value from adding generated MRI greater than using original BraTS data only, while italics indicates the opposite.*

| Data | | Brats | Brats +1/4 GAN | Brats + 1/2 GAN | Brats + 3/4 GAN | Brats + All GAN |
|---|---|---|---|---|---|---|
| DSC | ET | 0.559 | *0.506* | **0.607** | *0.520* | **0.607** |
|  | WT | 0.818 | *0.789* | 0.817 | **0.841** | 0.828 |
|  | TC | 0.648 | *0.638* | **0.701** | 0.664 | 0.683 |
| Sens. | ET | 0.704 | **0.795** | **0.843** | **0.782** | *0.621* |
|  | WT | 0.887 | **0.899** | 0.887 | *0.829* | *0.796* |
|  | TC | 0.662 | **0.740** | **0.751** | **0.769** | *0.648* |
| Spec. | ET | 0.985 | *0.985* | **0.989** | 0.985 | **0.990** |
|  | WT | 0.987 | *0.979* | 0.987 | **0.994** | **0.994** |
|  | TC | 0.992 | *0.990* | **0.994** | *0.991* | **0.995** |
| HD (mm) | ET | 11.8 | *16.2* | **11.2** | *13.7* | **8.5** |
|  | WT | 17.0 | *23.1* | 17.0 | **11.4** | **11.7** |
|  | TC | 17.4 | *22.2* | **17.0** | **16.8** | **13.4** |

Though we did not observe a strong correlation between the improvement of values of the metrics for each structure and amount of synthetic data added, there is a linear relationship between the amount of synthetic data added and the number of best overall validation metric value (underlined and bolded values). Adding a randomized quarter synthetic subset reports best overall sensitivity for WT, half synthetic subset gives best overall metrics in three categories, three-quarters shows best overall metrics in four areas and adding all synthetic data to double the training dataset size gives best overall metrics in 6 areas.

Training using all real images combine with half the synthetic images resulted in highest TC DSC value, while ET DSC value performed similarly to the model that used all synthetic images during training. This deserves a closer look as it seems to combat the hypothesis that the more synthetic images used, the better the model performs. However, the HD values, a metric some consider just as valuable as DSC, for ET/TC are much worse when using half the synthetic images than when training using all of them. This similarity in DSC and discrepancy in HD could come from using soft dice as the loss function.

The highest WT metrics (DSC, specificity, and HD) resulted from training on all real images and a three-quarter subset of the synthetic images. If only the WT requires segmentation, this model should be used although these values are comparable with validation results when using all synthetic images.

The synthetic MR images generated by the generative adversarial network showed increased performance in HD and DSC when used in totality during training. The difference in sensitivity and specificity can be explained by differing size of predicted contours.

**4.0 Discussion**

**4.1 GAN Benefits**

It has been reported in multiple studies that creating synthetic images using a generative adversarial network increases delineation accuracy. [17]This study investigated how the amount of GAN MR images included during the training process affects the U-Net's ability to accurately delineate the specified areas. Multiple U-Nets were trained using just BraTS data, BraTS and a randomized set of quarter, half, three quarters, and all GAN created images. To ensure similar level of quality images were used for all subgroups, the quarter randomized subset is included in the half synthetic MR image randomized subset, and the half synthetic MR image randomized subset is included in the three-quarter randomized subset. Obviously, all subsets were included in the total synthetic MR images dataset.

As expected, training on equal parts original real MR images and synthetic MR images increased validation results in most every metric over using real data only. Although, the WT contour metrics are not as high as using only three-quarters of the generated images, the increase in enhanced and tumor core metric should be noted. It can be observed that using GAN data combined with the original MR images from BraTS improves the results for DSC, specificity, and 95% CI HD. However, the sensitivity results decreased. We believe that the increased specificity and decreased sensitivity indicates the predicted validation contours created from the U-Net trained on both real and synthetic data differs in size from the ones trained from BraTS data only.

## 4.2 GAN Challenges

As stated in section 3.2 not all synthetic images possess the same quality. This could play a part in why adding only a randomized subset of a quarter of the generated images increased contour sensitivity but decreased DSC, specificity, and HD. Despite this studies effort, there is a possibility that the lack of obvious linearity between number of synthetic images used during training and validation results comes from the variation of image quality of synthetic MR images induced from the original image dataset. Future work includes training four U-Nets for the quarter subset and two for the half subset and taking the average results.

This lack of linearity shows it can be possible to achieve better validation results for certain contours without using all the synthetic or original MR images. To achieve the best results, it needs further investigation to understand the impact of the quality and distribution of the training data on the model performance.

## 4.3 INDIVIDUAL CASES

It is necessary to outline the best and worst cases to assess the model performance. To do this the two best performing and worst performing individual cases are presented. Figures 3-6 show whole tumor only for ease of viewing.

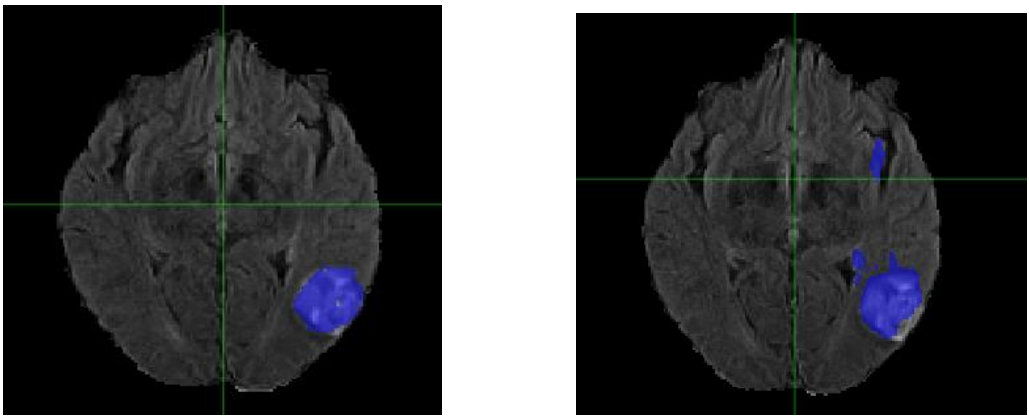

Figure 3: Case One (Improved). Flair MRI with contour from: both real and synthetic MRI (left) real MRI only (Right) DSC of ET/WT/TC was improved from 0.73/ 0.21/0.79 to 0.85/0.67/0.87. Blue is WT.

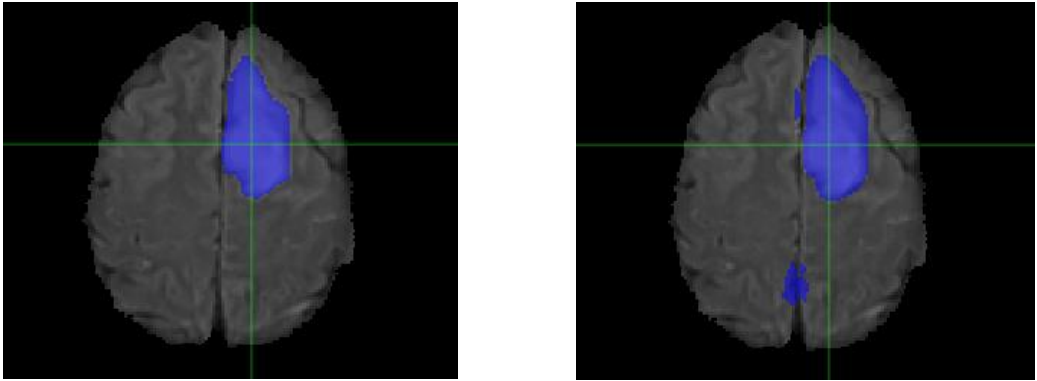

Figure 4: Case Two (Improved). Low-grade Glioma. Flair MRI with contour from: both real and synthetic MRI (left) real MRI only (Right) DSC was improved from 0/0.49/0.26 to 0/0.88/0.57. Blue is WT.

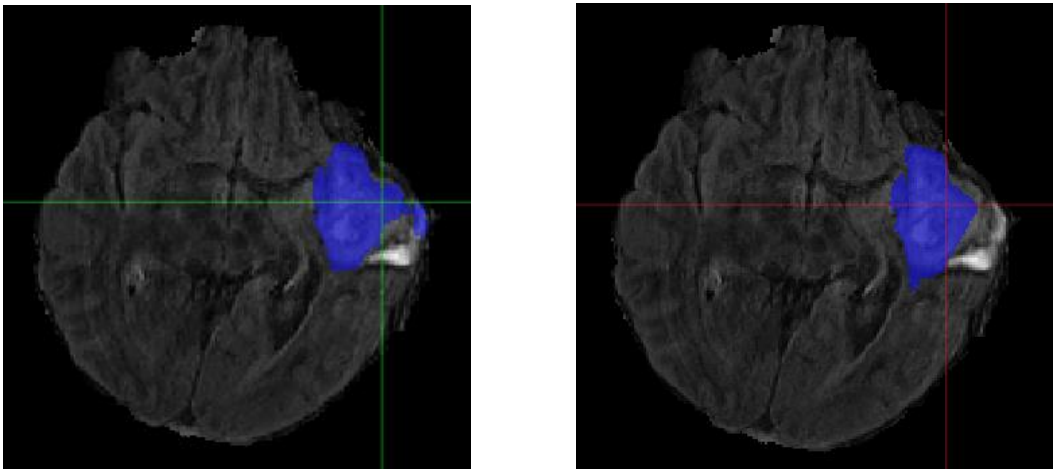

Figure 5: Case One (Worsened). Flair MRI with contour from: both real and synthetic MRI (left) real MRI only (Right) DSC of ET/WT/TC changed from 0.58/0.76/0.70 to 0.75/0.58/0.76. Blue is WT.

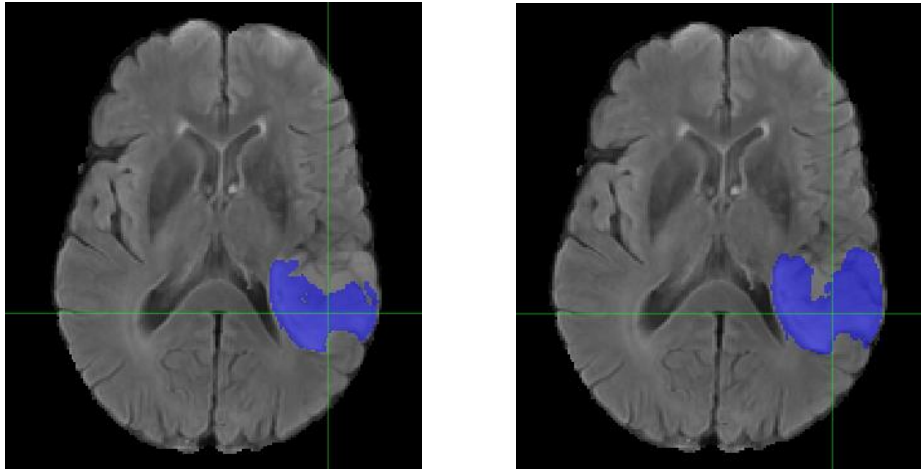

Figure 6: Case Two (Worsened). Low-Grade Glioma.  Flair MRI with contour from: both real and synthetic MRI (left) real MRI only (Right) DSC of ET/WT/TC  showed a decrease from 0.005/0.86/0.06 to 0.002/0.59/0.008. Blue is WT.

We have observed encouraging improvement of the segmentation accuracy for high grade glioma cases when the lesion was centrally and radially located. However, the challenge still exists in the low-grade glioma considering the difficulty to define the boundary. Location also plays a role in discerning whether the contouring accuracy would increase or decrease as the improved low-grade glioma was centrally located, while the one that showed worse results was located towards the edge of the brain.

## 5.0 Conclusion

We were able to generate high quality synthetic Flair, T2, T1, and T1CE MR images using the GAN and had a thorough evaluation of the images quantitatively and qualitatively.  The increased dataset showed promising results to improve the U-Net segmentation performance for overall DSC and HD. Such data augmentation strategy has the potential to address the small labeled data challenges in the medical image segmentation.

## 6.0 Author Contributions

NW designed the research and methodology; EC and ZD performed the data pre-processing, machine learning training and analysis, EC, NW, ZD, and JS wrote the paper; EL and JS provided clinical guidance on the evaluation of image quality and segmentation accuracy.

## 7.0 Acknowledgement

This work was supported by a Research Scholar Grant, RSG-15-137-01-CCE from the American Cancer Society.

[1] G. H. Ostrom QT, Truitt G, Boscia A, Kruchko C, Barnholtz-Sloan JS, "CBTRUS Statistical Report: Primary Brain and Other Central Nervous System Tumors Diagnosed in the United States in 2011-2015," *NCBI Neuro Oncol,* 2018.
[2] O. Ronneberger, P. Fischer, and T. Brox, "U-Net: Convolutional Networks for Biomedical Image Segmentation," *arXiv:1505.04597,*
[3] D. Shen, G. Wu, and H.-I.J.A.r.o.b.e. Suk, "Deep learning in medical image analysis," vol. 19, 221-248, 2017.
[4] I. Goodfellow, "Generative adversarial nets," *Advances in neural information processing systems,* 2014.
[5] M. a. S. J. a. p. a. O. Mirza, "Conditional generative adversarial nets," 2014.
[6] J. J. C. P. f. S. C. N. C. N. N. f. V. R. Gauthier, "Conditional generative adversarial nets for convolutional face generation," 2014.
[7] P. Isola, "Image-to-image translation with conditional adversarial networks," *Proceedings of the IEEE conference on computer vision and pattern recognition,* 2017.
[8] A. a. T. B. Dosovitskiy, "Generating images with perceptual similarity metrics based on deep networks," *Advances in neural information processing systems,* 2016.
[9] B. H. Menze *et al.*, "The Multimodal Brain Tumor Image Segmentation Benchmark (BRATS)," *IEEE Trans Med Imaging,* vol. 34, no. 10, pp. 1993-2024, Oct 2015.
[10] A. H. Bakas S, Sotiras A, Bilello M, Rozycki M, Kirby J, Freymann J, Farahani K, Da-vatzikos C. Segmentation Labels and Radiomic Features for the Pre-operative Scans of the TCGA-GBM collection
[11] A. H. Bakas S, Sotiras A, Bilello M, Rozycki M, Kirby J, Freymann J, Farahani K, Da-vatzikos C, "Segmentation Labels and Radiomic Features for the Pre-operative Scans of the TCGA-LGG collection," T. C. I. Archive, Ed., ed, 2017.
[12] S. Bakas *et al.*, "Advancing The Cancer Genome Atlas glioma MRI collections with expert segmentation labels and radiomic features," *Sci Data,* vol. 4, no. 170117, p. 170117, Sep 5 2017.
[13] A. Gomez-Iturriaga *et al.*, "Dose escalation to dominant intraprostatic lesions with MRI-transrectal ultrasound fusion High-Dose-Rate prostate brachytherapy. Prospective phase II trial," *Radiotherapy and oncology : journal of the European Society for Therapeutic Radiology and Oncology,* vol. 119, no. 1, pp. 91-6, Apr 2016.
[14] A. L. Maas, A. Y. Hannun, and A. Y. Ng, "Rectifier nonlinearities improve neural network acoustic models," in *Proc. icml*, 2013, vol. 30, no. 1, p. 3.
[15] K. Simonyan and A. J. a. p. a. Zisserman, "Very deep convolutional networks for large-scale image recognition," 2014.
[16] D. M. Pelt and J. A. Sethian, "A mixed-scale dense convolutional neural network for image analysis," *Proceedings of the National Academy of Sciences,* 10.1073/pnas.1715832114 vol. 115, no. 2, p. 254, 2018.
[17] J. D. Arnab Kumar Mondal, Christian Desrosiers, "Few-shot 3D Multi-modal Medical Image Segmentation using Generative Adversarial Learning," *arXiv,* 2018.